# Using Tracker to understand 'toss up' and free fall motion: a case study

Loo Kang WEE[1], Kim Kia TAN[2], Tze Kwang LEONG[3], Ching TAN[1]

[1] Ministry of Education, Educational Technology Division, Singapore
[2] Ministry of Education, Evergreen Secondary School, Singapore
[3] Ministry of Education, Raffles Girls School, Singapore

lawrence_wee@moe.gov.sg, tan_kim_kia@moe.edu.sg, tzekwang.leong@rgs.edu.sg, tan_ching@moe.gov.sg

Abstract:
This paper reports the use of Tracker as a computer learning tool in supporting effective learning and teaching of toss up and free fall motion for beginning secondary three (age 15 years old) students. This is a case study with (N=123) students of express-pure physics classes in a mainstream school in Singapore where we used a 8 multi-choice questions as a proxy to assess learning gains in pre and posttest to gauge the impact on learning. We found within experimental group gains with Cohen's effect size $d = 0.79 \pm 0.23$ (large effect) and normalized gains with a gradient of $<g>_{total} = 0.42 \pm 0.08$ (medium gain) above the traditional baseline value of $<g>_{non\ interactive}=0.23$ for all the 6 teachers, 3 classes of students who participated in this study.
Initial research findings suggest that allowing learners to relate abstract physics concepts to real life through coupling traditional video analysis and eventually video modeling could be an innovative and effective way to learn free fall motion.
2015 Resource: http://iwant2study.org/ospsg/index.php/interactive-resources/physics/02-newtonian-mechanics/01-kinematics/151-ball-toss-up

## I. INTRODUCTION

Many novice young students harbor misconceptions (Kavanagh & Sneider, 2006) about free fall motion and its scientific representation of the displacement versus time ($y$ vs $t$), velocity versus time ($v_y$ vs $t$), and acceleration versus time ($a_y$ vs $t$), graphs and when these representations are not well understood coherently, it is difficult to use them for deductive reasoning for different cases-scenarios like tossing a ball up with a higher initial speed or on the moon's surface for its velocity versus time graph.

While the use of real life examples, such as tossing a ball to demonstrate free fall can be performed by students in class, it is nevertheless challenging for students to translate the world view in $x$ and $y$ coordinates into typical scientific multiple representations (Wong, Sng, Ng, & Wee, 2011).

Our study involves the implementation of two 70 minutes computer laboratory lessons using worksheets with 3 pairs of teachers where each pair of teachers taught one class with a total number of students of N=123. A conceptual 8 item multi-choice questions was used in pre-posttests to serve as an indication of the learning gains after 3 weeks of total traditional and computer laboratory kinematics lessons.

Though not implemented in this study, the pedagogical approach called 'video modeling' (Brown, 2009) is further suggested for difficult to visualize velocity versus time graphs of two cases compare to tossing up a ball with a) with a greater force (higher initial velocity) on Earth and b) with the same force on Moon's surface. The free software tool Tracker (Brown, 2012a) can be downloaded from the Open Source Physics (Christian, Esquembre, & Barbato, 2011) website and has been used by authors in (Kinchin, 2012; Persson & Hagen, 2011; Poonyawatpornkul & Wattanakasiwich, 2013; Rodrigues & Carvalho, 2013; Wee, Chew, Goh, Tan, & Lee, 2012) Physics Education journal as well.

## II. INSTALLATION OF TRACKER

Tracker is a video analysis and modeling tool built on the Open Source Physics (OSP) Java framework. Though it is possible to run from a 5.3 Mb Tracker_486.jar file, we recommend using the respective installers found at http://www.cabrillo.edu/~dbrown/Tracker/, especially to enable the Xuggle video engine (Brown, 2012b) that can decode most video file formats. Installers for Tracker version 4.86 installers are available in Windows, Mac OS X as well as Linux operating systems.

## III. STUDY DESIGN

### A. Purpose of study

The study aims to determine the learning gains using within group Cohen's $d$ effect size and Hake's normalized (Hake, 1998) gain regression analysis from an infusion of two 75 minutes computer laboratory lessons (Figure 1) using the Tracker tool into existing typical Singapore school's teaching practices in the topic of kinematic of free fall.

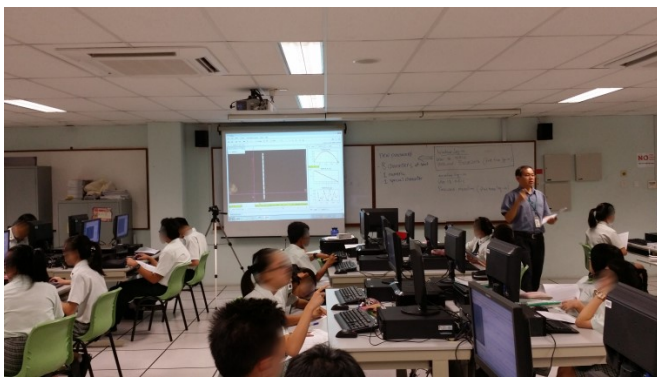

Figure 1. Typical computer laboratory lesson setup with teacher guiding about 40 students in a class in hands-on activity involving the use of Tracker to learn kinematics of free fall.

## B. Methodology

### 1) Research Design

A case study approach was adopted with the aim to provide a rich descriptive explanatory and analysis of a naturalistic social school setting to determine the effects of using Tracker on students' improved learning.

### 2) Particpants

Table 1. Class sizes of Experimental Group of the instructors. Students are taking pure Physics at the Ordinary Level.

| Teachers | Class | Number of students N |
|---|---|---|
| TKK and GCW | 3C | 41 |
| RT and NSH | 3I | 38 |
| ACS and SWL | 3R | 39 |
| Total | 3 | 123 |

The participants of this study are as shown in the Table 1, are facilitated by a pair of teachers in the 2x75 minutes computer laboratory lessons with students in the secondary three level typically in class size of about 40 in each class.

### 3) Lesson Plan

The lesson plan was to use these 2x75 minutes computer laboratory lessons to allow students to gain personal experiences in the physics of a toss-up of a ball motion. The worksheet serves as a guide to support the use of the tracker software and also to prompt students to predict, observe explain (Radovanović & Sliško, 2013) what they are supposed to make meaning of.

#### a) Teacher Professional Development

The 6 teachers were introduced to the use of Tracker about three months before the implementation of the lesson. Teacher TKK lead training sessions with the other secondary three physics teachers in the school as part of the Singapore school professional learning community-team effort, an initiative by the Ministry of Education Singapore in 2012.

They used the fortnightly regular meeting schedule to design the worksheet and Google site www.Tinyurl.com/evg3phy for professional development and maintaining consistent and high fidelity teaching practices.

#### b) Laboratory preparations

The computer laboratory technical assistant installed Tracker into the school's computers with one difficult encountered such as inability to launch Tracker program using the Windows|Start|Programs|Tracker|Tracker icon. The solution devised was to create a shortcut directly into the computer's Tracker installed folder such as C:\Program Files\Tracker\Tracker.jar as a workaround. Douglas has since fixed this error in subsequent releases of Tracker 4.86 as a result of our feedback.

#### c) Laboratory Activities

The teacher guide the students aided with a customized worksheet with the steps to open the software Tracker, load the video tossup.mov with the use of the teacher's computer and laboratory projector (Figure 1). The teacher typically show and explains how to use Tracker with one to two other teachers around to support the issues arising with the hands-on activities using computers.

## C. Data Collection Instruments

### 1) Pre-Post Test

An online[1] pre posttest was constructed referencing the activity and worksheets for aligning the learning tasks with the test items.

### 2) Focus Group Discussions with students

Post lessons focus group discussions were conducted with a total of nine students from teacher TKK class in three groups to further gather qualitative evidences of the lessons. The discussions also provided insights to data collected that the authors wanted to rationalize with such as the negative gains in pre-posttest registered by some students who scored 4-6 questions correct out of 8 (50-70%) in pre-test.

### 3) Informal discussions/ interviews with teachers

The discussions with the teachers suggested some of the students and teachers are not comfortable using computers and tracker software as it is new to them, they would require further support in conducting the lessons to high fidelity to the planned lessons. The pair of teachers serves to address that issue raised as the partner teacher would be able to co-teach and support the lesson with Tracker.

As this new worksheet was used for lessons for the first time, there were some difficulties that surfaced while being used by the students. The newly designed worksheet bears some of the benefits of our research such as being more appropriate scaffolds such as more targeted hints.

---

[1] https://docs.google.com/forms/d/1zm9NWWC7DWHkO2OYdbf150YyivJFPJeuT5vkO_KFKdI/viewform

Table 2: Percentage of students (total number of student, 123) who selected choices (a) – (d) on Problem Questions Q(1) to (8) on the Google Site. The correct response for each question has been italicized. The average correct score for pre-test is 33% and post-test is 48%

| Pre-test | | | | | | | | |
|---|---|---|---|---|---|---|---|---|
| Q | 1 | 2 | 3 | 4 | 5 | 6 | 7 | 8 |
| (a) | 18 | 34 | *39* | 46 | 79 | 4 | 27 | 15 |
| (b) | *55* | 29 | 24 | 31 | *17* | 13 | 27 | 18 |
| (c) | 25 | *34* | 30 | *11* | 2 | *24* | *40* | 26 |
| (d) | 2 | 3 | 8 | 12 | 2 | 58 | *6* | *41* |
| Post-test | | | | | | | | |
| Q | 1 | 2 | 3 | 4 | 5 | 6 | 7 | 8 |
| (a) | 11 | 22 | *57* | 51 | 42 | 7 | 23 | 19 |
| (b) | *71* | 20 | 14 | 22 | *50* | 14 | 28 | 34 |
| (c) | 17 | *56* | 26 | *21* | 6 | *54* | *44* | 16 |
| (d) | 1 | 2 | 3 | 6 | 2 | 25 | 4 | *31* |
| change | 16 | 23 | 19 | 10 | 33 | 30 | 4 | -10 |

*D. Data Collection Procedures*

At the beginning of the first computer based lessons, students were told to complete the pretest at the first 10-15 minutes of the lesson after they manage to login the school laboratory computers, navigate their internet browsers to the lesson Google site www.Tinyurl.com/evg3phy. Students were encouraged to complete the pre-test as it would give the teachers as idea of what parts to focus on later in class. Some students managed to discuss their answers thus it could have contributed to students with 50, 62.5 and 75% correct scores in pretest registering lower posttest scores later.

After the end of the topic on kinematics which is about 3 weeks after the first lesson, the students were brought back to complete the same items posttest individually without discussions.

*E. Results*

Table 2 is the results of the pre and posttest with 8 kinematics equations tabulated in percentage from the 123 students. The average correct score for pre-test is 33% and post-test is 48%.

*1) Question 1 to 6*

Question 1, 2 and 3 test students' ability to recall and identify simple $y$ versus time $t$, $v_y$ versus $t$ and $a_y$ versus $t$ graphs, each registering gains ranging from 16 to 23%.

Question 4 is a common conceptual test item that requires students to realize that the velocity and acceleration at the highest point of the vertically toss up motion is zero and non-zero respectively. The change is +10%

Question 5 and 6 are simple understanding test items requiring students to realize that when the ball is moving upwards and downwards, the gradient of the displacement-time graph is positive & decreasing in magnitude and negative & increasing in magnitude respectively. The change is higher at around 30%

*2) Question 7 and 8*

Question 7 and 8 are application test items where in Q7 the $v_y$ versus $t$ graph on Earth is given (Green), the higher initial velocity would result in a parallel but higher $v_y$ intercept line (Blue). The change is only 4% for Q7. As for Q8 the $v_y$ versus $t$ graph on Earth is given (Green), a lower gravitational acceleration would result in a line with the same intercept but a smaller in magnitude gradient. This time the change is – 10%.

*F. Discussions*

*1) Question 1 to 6*

Questions 1 to 6 registered positive gains which are to be expected. Interestingly, Question 4 option (a) also register a small +5% gains from 46% to 51% suggests that the misconception that object at the top of the motion continues to have zero acceleration, remains a difficult concept to understand and we hope in future to design better lessons to address this.

*2) Question 7 and 8*

Question 7 is roughly unchanged with +4% gain is not surprising as the teachers did not explicate this concept using Tracker's modeling pedagogy (Wee et al., 2012) but we argue holds great potential for experiential learning and deepening understanding.

Question 8 surprisingly registered -10 gains which after analyzing the option (b) suggests students are "tricked" by the effects of "due to lower mass with higher air resistance" not realizing it will be a terminal velocity trail when time is large whereas the correct answer of (d) is a linear trail suggesting constant acceleration with no drag.

*G. Results*

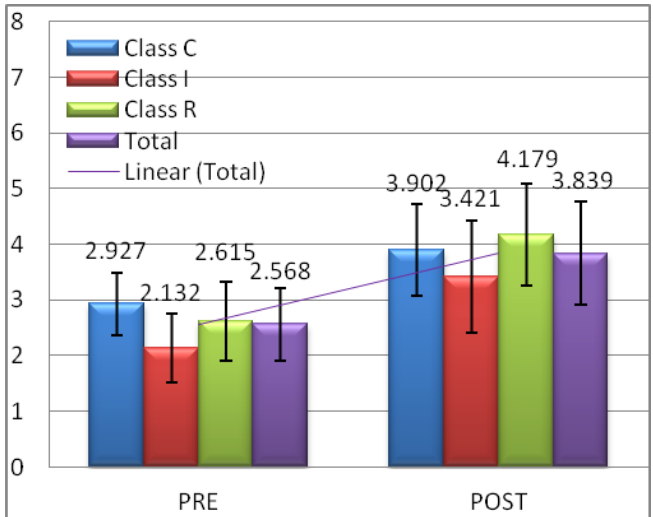

Figure 2. bar chart of pre-posttest scores of students from class C (BLUE), I (RED), R (GREEN) and the combined total (PURPLE) (left to right bars) showing total pre-test score and posttest score $C_{pre,post}$= (2.927 ± 1.237, 3.902 ± 2.022) , $I_{pre,post}$= (2.132 ± 1.104, 3.421 ± 1.648) , $R_{pre,post}$= (2.615 ± 1.407, 4.179 ± 1.824) and $TOTAL_{pre,post}$= (2.568 ± 1.298, 3.839 ± 1.868).

Based on the results of pre =1, post=2 test of eight multi-choice questions scores collected, all three classes' ($N_1 = N_2 = 123$) registered positive gains after infusion of two hands-on computers based laboratories with Tracker into traditional three week of lessons on the topic of kinematics of a toss-up free fall motion. All classes C (BLUE), I (RED), R (GREEN) registered gains (Figure 2) and worth noting is the total scores (PURPLE) are $\bar{x}_1 \pm s_1$ = (2.568 ± 1.298) and $\bar{x}_2 \pm s_2$ = (3.839 ± 1.868). This translates to Cohen's $d$ effect size = 0.79 ± 0.23 using equation (1) which it can be interpreted as large effect (Cohen, 1977) or practically significant (Wolf, 1986).

$$d = \frac{\bar{x}_2 - \bar{x}_1}{\sqrt{\frac{(N_1 - 1)s_1^2 + (N_2 - 1)s_2^2}{N_1 + N_2 - 2}}} \quad (1)$$

Using the normalized gain (Hake, 1998) $<g>$ in equation (2) where posttest scores and pretest scores are $x_2$ and $x_1$ respectively, 8 is the maximum score of the test, we analyzed the normalized learning gains in percentage across the three classes' pretest scores in percentage as the horizontal axis.

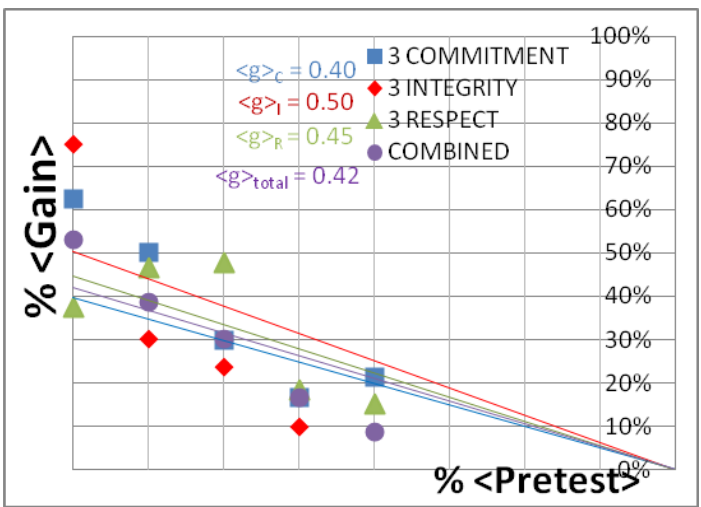

$$<g> = \frac{x_2 - x_1}{8 - x_1} \quad (2)$$

Figure 3. Gain vs Pretest of all three classes with pretest scores ranging from 0 marks, 0% to 4 marks, 50% used on the horizontal axis versus the gain in percentage. Notice the gains range from 0.40 to 0.50 with $<g>$total = 0.42 suggests the lessons are as effective as most interactive engagement lessons above the traditional lesson gains of 0.23 as reported in Hake, 1998a

We used only the pre-test data equal and below 50% as there are generally negative $<g>$ for pretest scores of 62.5% and 75% with no students with 87.5% or 100%. Our interviews with 8 students suggested that the negative gains could be contributed to a longer pre-test time 15 minutes compared to post-test 10 minutes and benefited from peers discussions during pre-test.

The general trend is not adversely affected by neglecting scores from 62.5 and 75%, thus to simplify the presentation, we choose scores 50% and below. The result of passing through (0%,0%) linear regression using data of $<g>$ in percentage versus pretest scores in percentage (Figure 3) suggests, the three classes normalized gains $<g>$ are near the medium gains $<g>_C$=0.40 (BLUE), $<g>_I$=0.50 (RED), $<g>_R$=0.45 (GREEN) and $<g>_{total}$ = 0.42 (PURPLE) in the range of gradients well above the traditional normalized gain of $<g>_{traditional}$=0.23.

Based on standardized mean difference Cohen's d = 0.79 (large effect) and normalized gain $<g>$=0.42 (medium gain) analysis, the evidences suggest the students did learn the kinematics concepts better than traditional passive non-interactive lessons.

We recommend design-based research (Juuti & Lavonen, 2006) method to continually improve this Tracker based lessons as this is the first year the teachers have conducted this lessons as such we believe there is still a lot of room for improving the implementation of these lessons, for deepening learning experiences through video modeling (Brown, 2012a) elaborated later.

IV. STUDENTS' REFLECTION ON TRACKER LESSON

To give some themes into the conditions and processes during the laboratory lessons, the following are some excerpts from the informal interviews with the students. Words in brackets [] are added to improve the readability of the qualitative interviews.

*1) Tracker supports interactive and real world physics*

"We are able to see the connections between the real life [video] and the [scientific] graph[s]. Tracker helps me to confirm the theory [in kinematics] I have learned."

"The video analysis [Tracker] gives me the opportunity to check the data collected. I realized that in real life data collection, there are random errors, which was shown from the graph plotted."

"Compared to teachers' explanation on the board, the video analysis gives us more opportunity to have the real learning experience; rather that spoon feed us with content. By allowing us to use the video analysis [tool, Tracker], we are able to see more precisely between the ball and the graph plotted."

"Perhaps we can have a practical lesson [a performance task] in the curriculum. We would be interested in trying it out ourselves to do the experiment and record the videos ourselves".

*2) Overcoming initial difficulties using Tracker*

"I don't have the experience to load the video and track the video. I would like teachers to use the video Tracker to show us the scenarios in learning, so that we can strike a balance between learning effectively and not to spend too much time in setting up the video [tool] Tracker".

To address this difficult, teacher TKK has YouTube video tutorials found http://www.youtube.com/user/kimkiatan/videos to help his students.

## V. TEACHERS' REFLECTION

Despite the 'relative' success with Cohen's *d* effect size = 0.79 $\pm$ 0.23 and normalized gain $<g>_{total}$ = 0.42 in the learning using tracker, the teachers reflected and 3 recommendations are derived as areas of improvement.

*A. Start using Tracker for easier horizontal kinematics task*

Start the year with an easier horizontal kinematics task like investigating a constant speed object moving on a frictionless track. This serves to address the cognitive overload (Roth, 1999) problem encountered when using the software Tracker and the relatively complex concept of toss up and free fall for fresh secondary three students.

*B. Practice of Video Modeling especially for Question 7 and 8*

As mentioned before, the teachers were only aware of the video modeling approach later and unable to enact the learning with this approach across all the 3 classes. Thus, subsequent teaching interventions will include teacher's direct instruction of the video modeling activities not explained in this paper.

*C. Integration of tracker use in topics on kinematics, dynamics and work, energy*

We also recognize for sustained learning gains that Tracker use be integrated for topics like kinematics, dynamics and work, energy which tracker's analysis affords for seamlessly.

## VI. CONCLUSION

This is a case study with (N=123) students of express-pure physics classes in a mainstream school in Singapore where we used a 8 multi-choice questions as a proxy to assess learning gains in pre and posttest to gauge the impact on learning. We found the within experimental group gains of effect size of 0.79 $\pm$ 0.23 (large effect) and normalized gains within the expected linear regression with a gradient of $<g>_{total}$ = 0.42 (medium gain) for all the 6 teachers and 3 classes in this study. The evidences suggest the students did learn the kinematics concepts better than traditional passive non-interactive lessons.

We make 3 recommendations to further strength learning by a) start using Tracker for easier horizontal kinematics task to lower cognitive loading for novice students b) using the video modeling pedagogical approach (Brown, 2012a; Wee et al., 2012) to improve learning targeted at Question 7 and 8 and c) integrate tracker use in topics on kinematics, dynamics and work & energy for sustainable and seamless learning with the tool Tracker.

### ACKNOWLEDGMENT

We wish to acknowledge the passionate contributions of Douglas Brown, Wolfgang Christian, Mario Belloni, Anne Cox, Francisco Esquembre, Harvey Gould, Bill Junkin, Aaron Titus and Jan Tobochnik for their creation of Tracker video analysis and modeling tool.

This research is made possible thanks to the eduLab project NRF2013-EDU001-EL017 "Becoming scientists through video analysis" awarded by the National Research Foundation (NRF), Singapore in collaboration with National Institute of Education (NIE), Singapore and the Ministry of Education (MOE), Singapore.

AUTHOR

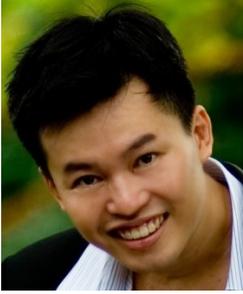 Loo Kang WEE Lawrence is currently educational technology specialist II at the Ministry of Education, Singapore. His open source physics contribution garner awards including Public Service PS21 Distinguished Star Service Award 2014 and Best Ideator 2012, Ministry of Education, Best Innovator Award 2013 and Excellence Service Award 2012.

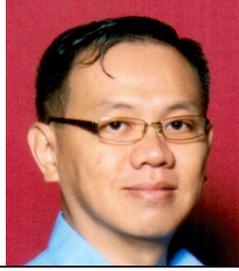 Kim Kia TAN is a Physics teacher at Evergreen Secondary School, Singapore. He currently leads a Professional Learning Team in teaching and learning of Physics using ICT in his school.

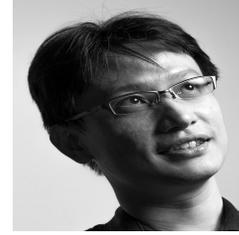 Tze Kwang LEONG is currently teaching physics in Raffles Girls' School. His interest is in demonstrations and ICT tools that augment lessons to deepen student understanding.

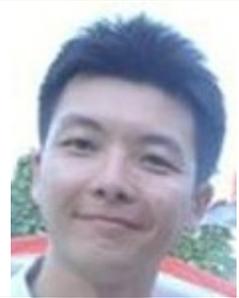 Ching TAN is an Educational Technology Officer from Educational Technology Division, Ministry of Education, Singapore. He is a science teacher with a keen interest in visual thinking and multiple representations.